\documentstyle[12pt]{article}
\newcommand{\beq}{\begin{equation}}
\newcommand{\eeq}{\end{equation}}
\newcommand{\beqar}{\begin{eqnarray}}
\newcommand{\eeqar}{\end{eqnarray}}

\newcommand{\bd}{\begin{itemize}} 
\newcommand{\ed}{\end{itemize}} 
\newcommand{\bc}{\begin{center}}
\newcommand{\ec}{\end{center}}

\newcommand{\be}{\begin{equation}}
\newcommand{\ee}{\end{equation}}
\newcommand{\ba}{\begin{array}}
\newcommand{\ea}{\end{array}}

\begin{document}


\begin{center}
{\large
{\bf Chaotic dynamics in the framework of the Vlasov-Nordheim equation}}
\\[6ex]
{ M. Baldo, G.F. Burgio and A. Rapisarda{\footnote{{\small
Talk given by A. Rapisarda.\\
\noindent}}} } \\[4ex]

{\sl{  Universit\'a di Catania and INFN Sezione di Catania}}

{\sl {  Corso Italia 57, I-95129 Catania, Italy }}\\[5ex]

{ABSTRACT}

\end{center}

{\small {
In the framework of the time-dependent mean field theory, we solve the
Vlasov-Nordheim equation and study the sensitive dependence of this highly
non-linear equation on the initial conditions. We find that,  when nuclear
matter is inside the spinodal region, initially small differences between
similar trajectories grow  exponentially and produce strongly divergent
paths resulting in a chaotic evolution. We focus on the fact that chaos
spontaneously arises without putting any additional noise. We calculate the
largest Lyapunov exponent for various initial densities. This analysis is
performed for nuclear matter in two dimensions. It is argued that the mean
field chaoticity can cause nuclear multifragmentation
observed in intermediate energy heavy ion reactions.}}\\[4ex]

{\section {\bf Introduction}}

Intermediate energy heavy ion collisions  offer a unique possibility to
study nuclear matter under extreme  conditions of density and temperature.
In the energy range between 20 and 100 MeV/nucleon, the formation of  hot
nuclei and their successive decay, through emission of several
intermediate mass fragments (IMF : Z $\geq$ 2) with large fluctuations
from event to event, has stimulated a huge debate on the main mechanism
responsible of this phenomenon, commonly named "multifragmentation".

Both statistical models  and dynamical simulations \cite{gro82,bau92} have
been widely used providing suggestions on the possible paths followed by
the nuclear system along the multifragmentation process. Surface
instabilities or ring formation \cite{bau92} have been recently considered
as possible steps towards fragment formation. Quite surprisingly also the
percolation model has been successfully used  \cite{plo90}, suggesting a
universal behavior typical of phase transitions and common to different
fields. The onset of instability in the spinodal zone of the nuclear matter
Equation of State (EOS) has been studied in ref. \cite{hei93}, by analyzing
the corresponding linear response for imaginary frequency as a function of
the wave number and of the density, even including a noise of Langevin-type
to the collision integral in  the BUU equation \cite{bur92}. However
already in the 80s, in one of the  first attempts to  describe
multifragmentation Knoll {\it et al.} \cite{kno84} performed numerical
calculations in the framework of  two-dimensional Time Dependent Hartree
Fock (TDHF) equation which showed many of the features described by the
recent BUU  calculations \cite{bau92}. In particular by evolving an
ensemble of Slater determinants, they found a dependence on the  initial
conditions and a spontaneous symmetry breaking which lead to large
fluctuation from event to event and in the charge or mass distributions.
This scenario suggests a strong non-linear evolution, as recently
suggested also in ref.\cite{idi93,bbr}.

In order to clarify the mechanism of multifragmentation, we study in this
contribution the dynamics of the Vlasov-Nordheim (VN) equation in  the
spinodal zone. We find unambiguously a strong non-linear evolution which
implies a sensitive dependence on the initial conditions and a  spontaneous
symmetry breaking \cite{bbr}. This means that the dynamics in the  spinodal
region is chaotic, although the equation is deterministic: similar
trajectories can have a completely different evolution once they enter the
spinodal region. We estimate also a Lyapunov exponent, which gives the mean
rate of  divergence between the trajectories in density space, in line with
the usual analysis of dynamical systems \cite{ben76}.   We find that
two-body collisions do not affect  the onset of chaoticity in nuclear
many-body systems. Finally we draw  some conclusions. Further details can
be found in ref. \cite{bbr}.
\\[3ex]

{\section {Theoretical framework}}

{\subsection {\sl {The Vlasov-Nordheim equation}}}

The numerical analysis is performed by solving the Vlasov-Nordheim
equation
\be
\label {1}
{\partial f \over \partial t} - \{H[f], f\} = I[f]
\ee
on a two-dimensional lattice, using the same code of ref.\cite{bur92}. In
eq. (1) $H[f]$ is the self-consistent effective one-body  Hamiltonian,
while $I[f]$ is the two-body Pauli blocked collision integral. The single
particle phase space is divided into cells and nuclear matter  is confined
in a large torus with periodic boundary conditions. The side lengths of the
torus are equal to $L_x=51~fm$ and  $L_y=15~fm$. In order to achieve
sufficient accuracy, very small cells having $\delta x = {1\over3}$ fm and
$\delta p=40$ MeV/c  are employed. The time integration step was chosen
equal to 0.5 fm/c. We adopt a simplified Skyrme interaction for the
effective one-body field which is averaged over the y-direction
\cite{bur92} and convoluted over a Gaussian function whose width is $\mu =
0.43$ fm. This is equivalent to introduce a finite effective range. We
employ a Fermi momentum at saturation of $P_F=260$ MeV/c, which gives a
density in two dimensions equal to $\rho_0=0.55 ~fm^{-2}$.
\\[2ex]

{\subsection {\sl{ Definition of the largest Lyapunov exponent}}}

In the theory of dynamical systems, a measure of chaoticity is expressed in
terms of Lyapunov exponents \cite{ben76}.  The largest Lyapunov exponent
$\lambda_{\infty}$ can be extracted by considering at a given time $t$  the
distance $d(t)$  between two trajectories, along an unstable direction in
phase space,  and define
\be
\lambda(t)\, =\, {\log (d(t)/d_0) \over t}  ~,
\ee

where $d_0 = d(0)$. The average of this quantity along the trajectory
gives the mean rate of exponential separation of neighbouring
trajectories
\be
\lambda_{\infty} \, = \, \lim_{t\to\infty}\, \lim_{d_0\to0}\, \lambda(t)
\ee

In numerical applications one has to select a series of small values of
$d_0$, and check that the corresponding values of $\lambda(t)$  stabilize
around a definite value of $\overline{\lambda}$.

In order to characterize the system as a whole we have chosen as distance
between two trajectories the difference in norm between their density
profiles
\be
d(t) \, =\, \sum_i \, | \rho^{(1)}_i(t) - \rho^{(2)}_i(t) |/N_c ~,
\ee

where the index $i$ runs over the $N_c$ cells in ordinary space, and
$\rho^{(1)}_i$, $\rho^{(2)}_i$ are the densities in the cell $i$ for the
trajectories $1$ and $2$ respectively. The definition of eq. (4) is
sufficient  for the present analysis, although possible differences in
momentum space  are averaged out. It should include the contribution of all
the unstable modes which dynamically grow up during the evolution.
\\[3ex]

{\section {Results and discussion}}

First we analyze the mean field dynamics neglecting two-body collisions,
i.e. solving the left-hand side of eq.(1). We have initialized nuclear
matter with a   sinusoidal profile along the x-direction, characterized by
a wave number  $k = 2\pi (n_k/L_x)$ with $ n_k = 5$ and a small amplitude
equal to 1$\%$ of  the local density. The local momentum distribution is
assumed to be the one of a Fermi gas at a temperature $T =~3~MeV$.

In fig.1a we show the evolution of the density profile for different times
at normal density. The initial oscillation along the x coordinate is
rapidly  damped and,  as fig.1b shows,  the evolution of the Fourier
transform of the density profile presents only one  peak which disappears
during the same time scale. Changing the initial density by a small amount,
for example $d_0={\Delta\rho\over\rho_0}=10^{-2}$, one obtains exactly the
same behavior and the  evolution of the two events is  indistinguishable.
This is typical of the "regular" region of a dynamical system, and
indicates  the stability of the dynamics with respect to small
perturbations.

On the contrary, if the  density $\rho$ is chosen well inside the spinodal
region, $\rho = 0.5 \rho_0$ in fig.2, the initial symmetry of the density
profile, after a short time, is completely broken and fluctuations are
rapidly enhanced.


The Fourier transform on the other hand shows the appearance of several new
modes which where not excited initially. Moreover a small change in the
initial density, see fig.2bis where $\rho = 0.51 \rho_0 $, produces an
evolution which is completely different  both in shape and in amplitude
from that of fig.2.  This is the typical behavior of a dynamical system in
a chaotic  region, where due to the non-linearity of the equation, very
small initial  perturbations are rapidly amplified and  the distance
between similar trajectories diverges exponentially.

In our case this behavior occurs even if
$d_0={\Delta\rho\over\rho_0}=10^{-4}$,  see fig.4. In general one finds an
increase of the populated wave numbers together with a strong mixing as the
evolution proceeds, despite the fact that only one wave number $k$ is
initially occupied. The final frequencies spectrum has little resemblance
with the initial one. Note however that the  initial mode is not completely
damped, and some memory of it is still retained. We have checked that this
general trend depends neither on the initial $n_k$ nor on the shape of the
initial profile.

In fig.3a it is displayed the time evolution of the distance between
trajectories, as defined in eq.(4), normalized with respect to its initial
value $d_0 =~10^{-4}$. We note that, while the distance  remains constant
in the regular zone as time goes on, it increases exponentially when the
system starts in the spinodal region. In fig.3b $\lambda (t)$ as defined in
eq.(2) is plotted vs. time. As one can see, $\lambda$ tends to zero when
dynamics is regular, indicating that the trajectories actually do  not
diverge between each other. On the contrary, $\lambda$ rapidly converges to
a value $\overline{\lambda}$ which is not zero in the unstable regions and
unambiguously defines the largest Lyapunov exponent $\overline {\lambda}$.
We note that our time scale for the onset of chaos is in agreement  with
other numerical simulations on instability growth  \cite{idi93,kno84,bau92}
and with recent experimental results on multifragmentation \cite{bau93}.


We have checked that the results are stable with respect to the increase of
the numerical precision. This is shown in fig.4, where the time evolution
of $\lambda$ is plotted for different values of the time integration step
and initial distance. No appreciable variation of the results is obtained
using a value of $d_0$ smaller than $10^{-4}$ or  decreasing the time step.
Therefore the divergence of the trajectories is an intrinsic dynamical
property of the nuclear mean field in this region.

The inclusion of two-body collisions, performed by solving the VN equation
as done in ref.\cite{bur92}, but  without the Langevin term, does not
modify substantially the dynamics, as also found independently in
ref.\cite{idi93}.  This can be seen in fig.5, where $\lambda$ is displayed
vs. time for a density well inside the spinodal region. We note only a
slight variation - about 10 $\%$- of $\overline {\lambda}$ when  including
the collision term. In fact, since the dynamics is chaotic, the exponential
divergence between trajectories can hardly be affected by the collisions,
which have a small effect  at low density because of the diluteness of the
gas \cite{bur92,idi93}.


Finally in fig.6 we show the extracted values of $\overline{\lambda}$
calculated at different densities. The values of $\overline{\lambda}$
increase from zero at $\rho = 0.66 \rho_0$ to a  a flat maximum around the
density $\rho = 0.35\rho_0$.  The curve reported in fig.6 , which is
essentially independent from the initial $k$, clearly defines the nuclear
matter spinodal region up to its upper limit.

It is worthwhile to compare the characteristic times  $\tau_{MF} =
\hbar/\overline{\lambda}$,  which defines the time scale of the divergence
between mean  field trajectories, with the single particle characteristic
time  $\tau_{sp}=\hbar/E_F$, being $E_F$ the Fermi energy at the given
density. We have found that, for densities $\rho \leq 0.4\rho_0$, the
divergence time is smaller than the single particle time, and  therefore in
this region the notion itself of mean field ceases to  have any validity.
\\[3ex]

{\section {Conclusions}}

In the framework of a purely mean field theory, we have shown that the
nuclear dynamics inside the EOS spinodal region is chaotic. Without adding
any noise, and even without two-body collisions, spontaneous symmetry
breaking occurs leading to large unpredictable density fluctuations. We
characterize the chaoticity of the system through the largest Lyapunov
exponent, for which we suggest a computational recipe. The latter does not
depend on the particular choice of the initial conditions or the time
integration step, but it is strongly related to the initial density.  The
small influence due to two-body collisions has been discussed, as well as
the validity limits of the mean field theory. These results confirm
quantitatively a scenario already suggested by other authors
\cite{kno84,bur92,idi93,gro93} and can have important consequences for the
whole picture of nuclear multifragmentation.

Generally speaking, a central collision between heavy ions at intermediate
energy can be viewed as follows. In the first stage of the reaction nuclear
matter gets compressed and a highly excited composite nuclear system is
formed.  During this phase, the dynamics is insensitive to small
modifications of the initial conditions. Once the maximum compression is
reached, nuclear matter  starts to expand and can merge into the spinodal
region. According to the results presented in this paper, at this  point
the  dynamics becomes chaotic and  the corresponding large density
fluctuations dominate the cluster  formation. The degree of chaoticity and
fluctuations depends of course on the details of the dynamics which brings
the reaction inside the spinodal region. We would like to stress that,
since initially very close  events follow completely different dynamical
paths, an event-by-event  description is strongly advised. Averaging over
different theoretical paths can partially mask the non-linear evolution
\cite{chom93}. This scenario can support the behavior suggested by
percolation model and at the same time disagree with an equilibrium
statistical  description since some memory of the initial mode seems to
remain.  Unfortunately, at the moment, the extension of these calculations
to the case of collisions between heavy ions appears  problematic even in
two dimensions. A realistic theoretical simulation which is able to
describe how the system enters into the spinodal region  and how long it
stays there is desirable. On the other hand  an unambiguous experimental
signature is needed.\\[3ex]

\newpage
{\bf Figure captions}

\vspace {1 cm}
\noindent
{Fig.1 ~~~ Evolution of the den\-si\-ty pro\-fi\-le (a) and
its Fourier tran\-sform (b) at initial
normal density, see text.}
\vspace  {1 cm}

\par
\noindent
{Fig.2 ~~~ Evolution of the den\-si\-ty pro\-fi\-le (a) and
its Fourier tran\-sform (b) at initial
density ${\rho}=0.5 \rho_0$, see text.}
\vspace {1 cm}

\par
\noindent
{Fig.2bis ~~~ Evolution of the den\-si\-ty pro\-fi\-le (a) and
its Fourier tran\-sform (b) at initial
density $\rho=0.51 \rho_0$, see text.}
\vspace  {1 cm}

\par
\noindent
{Fig.3 ~~~ For three values of in\-it\-ial den\-si\-ty
the evolution of the di\-stan\-ce bet\-ween two trajectories
(a) and the corresponding $\lambda(t)$ (b) are shown.
The initial distance is $d_0=10^{-4}$, see text.}
\vspace {1 cm}

\par
\noindent
{Fig.4 ~~~ The quantity $\lambda(t)$ as a func\-tion of time
changing $d_0$ and the time step. Using a va\-lue of
$d_0$ smaller than $10^{-4}$ no change occurs.}
\vspace  {1 cm}

\par
\noindent
{Fig.5 ~~~ Behavior of $~\lambda(t)~$ as a function of time
with and without collision term in eq.(1).}
\vspace {1 cm}

\par
\noindent
{Fig.6 ~~~ The Lyapunov exponent $\overline\lambda$ vs. $\rho/\rho_0$
at temperature T=3 MeV, see text.}
\vspace  {1 cm}

\end{document}